\documentclass[cameraready]{Interspeech}

\usepackage{graphicx}
\usepackage{booktabs}   
\usepackage{multirow}
\usepackage{url}
\usepackage{amsmath}
\usepackage{placeins}

\title{Toward Multimodal Industrial Fault Analysis: A Single-Speed Chain Conveyor Dataset with Audio and Vibration Signals}

\author[affiliation={1},equalcontribution]{Zhang}{Chen}
\author[affiliation={1,4},equalcontribution]{Yucong}{Zhang}
\author[affiliation={1}]{Xiaoxiao}{Miao}
\author[affiliation={1,2,3},correspondingauthor]{Ming}{Li}
\address{
  $^1$ Digital Innovation Research Center, Duke Kunshan University, Kunshan, China \\
  $^2$ School of Artificial Intelligence, The Chinese University of Hong Kong, Shenzhen, China \\
  $^3$ School of Artificial Intelligence, Wuhan University, Wuhan, China \\
  $^4$ School of Computer Science, Wuhan University, Wuhan, China
}
\email{zc199@duke.edu,ming.li.cuhksz@gmail.com}


\begin{document}
\maketitle

\begin{abstract}
We introduce a multimodal industrial fault analysis dataset collected from a single-speed chain conveyor (SSCC) system, targeting system-level fault detection in production lines. The dataset consists of multimodal signals, including three audio and four vibration channels. It covers normal operation and four representative fault types under multiple speeds, loads, and both clean and realistic factory-noise conditions reproduced on-site. It is explicitly designed to support channel-wise analysis and multimodal fusion research. We establish standardized evaluation protocols for unsupervised fault detection with normal-only training and supervised fault classification with balanced dataset splits across different operating conditions and fault types. A unified channel-wise kNN baseline is provided to enable fair comparison of representation quality without task-specific training. The dataset offers a practical and extensible benchmark for robust multimodal industrial fault analysis.
\end{abstract}


\vspace{-2mm}\section{Introduction}\vspace{-2mm}

Early fault analysis is a critical component of industrial condition monitoring, as it enables timely maintenance and prevents unexpected system failures in automated production lines~\cite{garcia2025conditionreview}. With recent advances in signal processing and machine learning, fault analysis systems have increasingly adopted data-driven approaches, whose effectiveness strongly depends on the availability of high-quality and representative datasets. 

Despite growing interest in data-driven industrial condition monitoring, most publicly available fault analysis datasets are collected under laboratory conditions and focus on a limited set of machine types, such as electric motors or rolling bearings. Moreover, many existing datasets rely on a single sensing modality, typically audio or vibration signals, which may fail to capture complementary fault-related information available across multi-modality sensors. Noise conditions are often excluded entirely or introduced only through post-hoc synthetic data augmentation, resulting in a mismatch between benchmark evaluations and practical deployment scenarios where strong and structured environmental noise is inevitably present.

To address these limitations, we introduce a new multimodal industrial fault analysis dataset collected from a production-line–specific SSCC, capturing the behavior of a full system during the operation. The new SSCC dataset comprises synchronized multi-channel audio and vibration recordings acquired from heterogeneous devices, covering normal operation and multiple fault types under different operating conditions. While data collection is conducted in a laboratory setting, realistic deployment challenges are incorporated through controlled reproduction of factory noise recorded on-site, enabling systematic and reproducible evaluation under noisy conditions which have attracted growing interest in industrial settings~\cite{2023Moysidisnoiseeffect}. In addition, we establish unified evaluation protocols supporting both unsupervised fault detection and supervised fault classification tasks. Together, the proposed SSCC dataset and protocols provide a practical benchmark for evaluating multimodal signal representations and fault analysis methods in industrial condition monitoring. The code\footnote{Codes available at \url{https://github.com/Anonymous2417/SSCC-Fault-Benchmark.git}.}, and the dataset with demo\footnote{Dataset and demo available at \url{https://anonymous2417.github.io/SSCC-Dataset}.} are publicly available.

\vspace{-2mm}\section{Related Work}\vspace{-2mm}

Classical benchmarks such as the Case Western Reserve University~(CWRU)~\cite{cwru} bearing fault dataset, the IDMT-ISA Electric Engine~(IIEE)~\cite{iiee} dataset, and the IDMT-ISA Compressed Air~(IICA)~\cite{iica} dataset are primarily designed for vibration-based fault analysis and are collected under controlled laboratory conditions. These datasets have played an important role in advancing data-driven methods, but they typically focus on a limited set of machine components, such as bearings or motors, and rely on a single sensing modality.

To our knowledge, MaFaulDa~\cite{mafaulda} and the HUSTmotor multimodal dataset~\cite{zhao2025multimodal} are among the few publicly available datasets that provide both vibration and audio signals for industrial fault analysis. Both datasets are based on rotating machinery platforms and employ a single audio channel alongside multi-axis vibration measurements. MaFaulDa is collected under relatively clean laboratory environments without explicit environmental noise modeling, while HUSTmotor introduces noise primarily through additive signal-level processing rather than through real acoustic interference recorded from factory settings.

In contrast, the SSCC dataset is constructed on a SSCC system and incorporates three different audio recording devices in parallel. Besides, it integrates real factory environmental noise into the data collection process. This design facilitates more realistic evaluation of multimodal fault analysis under practical industrial acoustic conditions.

Different from datasets that focus on individual components, the SSCC dataset targets a complete conveyor system. It adopts a system-level fault analysis perspective. Multiple sensors are placed at different locations across the machine. This setup better reflects real industrial deployment, where faults can propagate across coupled components. The dataset supports both fault detection and fault classification tasks under unified evaluation protocols, and facilitates research on multimodal learning with richer channel diversity in industrial fault analysis.

\begin{table*}[!t]
\centering
\caption{Overview of the SSCC dataset. Rows represent fault categories~(defined in Section~\ref{subsec:fault_type}) and load conditions, while columns correspond to conveying velocities under clean and environmental noise settings. Each cell reports the number of samples, and “--” indicates unavailable conditions. For low speeds (20 and 40), data are collected only under the normal clean setting, as such speeds are uncommon in realistic industrial operation.}\vspace{-3mm}
\label{tab:dataset_overview}
\resizebox{.8\textwidth}{!}{\large
\begin{tabular}{llcccccccccc|c}
\toprule
\textbf{Category} & \textbf{Load}
& \multicolumn{2}{c}{\textbf{Velocity=20}}
& \multicolumn{2}{c}{\textbf{Velocity=40}}
& \multicolumn{2}{c}{\textbf{Velocity=60}}
& \multicolumn{2}{c}{\textbf{Velocity=80}}
& \multicolumn{2}{c}{\textbf{Velocity=100}}
& \textbf{Total} \\
& & \textbf{clean} & \textbf{noise}
  & \textbf{clean} & \textbf{noise}
  & \textbf{clean} & \textbf{noise}
  & \textbf{clean} & \textbf{noise}
  & \textbf{clean} & \textbf{noise}
  & \\
\midrule
normal & heavy & 69 & -- & 68 & -- & 69 & 96 & 77 & 97 & 68 & 96 & 640 \\
normal & med   & 73 & -- & 66 & -- & 78 & 96 & 67 & 94 & 66 & 99 & 639 \\
normal & light & 68 & -- & 86 & -- & 69 & 97 & 66 & 97 & 67 & 97 & 647 \\
\midrule
dry & heavy & -- & -- & -- & -- & 57 & 57 & 57 & 57 & 56 & 58 & 342 \\
dry & med   & -- & -- & -- & -- & 69 & 57 & 56 & 57 & 64 & 56 & 359 \\
dry & light & -- & -- & -- & -- & 62 & 58 & 49 & 56 & 57 & 57 & 339 \\
\midrule
lean & heavy & -- & -- & -- & -- & 57 & 57 & 57 & 57 & 57 & 57 & 342 \\
lean & med   & -- & -- & -- & -- & 58 & 56 & 57 & 57 & 56 & 55 & 339 \\
lean & light & -- & -- & -- & -- & 57 & 56 & 69 & 60 & 56 & 56 & 354 \\
\midrule
loose & heavy & -- & -- & -- & -- & 96 & 99 & 97 & 97 & 97 & 96 & 582 \\
loose & med   & -- & -- & -- & -- & 73 & 96 & 96 & 96 & 96 & 99 & 556 \\
loose & light & -- & -- & -- & -- & 96 & 96 & 96 & 96 & 96 & 97 & 577 \\
\midrule
screwdrop & heavy & -- & -- & -- & -- & 57 & 56 & 57 & 55 & 56 & 54 & 335 \\
screwdrop & med   & -- & -- & -- & -- & 57 & 47 & 52 & 51 & 87 & 54 & 348 \\
screwdrop & light & -- & -- & -- & -- & 56 & 56 & 56 & 56 & 45 & 56 & 325 \\
\midrule
\textbf{Overall} & \textbf{Total}
& \textbf{210} & --
& \textbf{220} & --
& \textbf{1,011} & \textbf{1,080}
& \textbf{1,009} & \textbf{1,083}
& \textbf{1,024} & \textbf{1,032}
& \textbf{6,669} \\
\bottomrule
\end{tabular}
}\vspace{-5mm}
\end{table*}

\vspace{-4mm}\section{Dataset Overview}

\vspace{-1mm}\subsection{Machine System Description}\vspace{-1mm}

The proposed SSCC dataset is collected from a single-speed chain conveyer system, a production-line--specific mechanical structure widely used in industrial conveying and transmission applications. Unlike laboratory test rigs that focus on isolated components such as bearings or motors, the considered system represents a complete chain-driven conveying mechanism commonly deployed in automated production lines.

\begin{figure}[t]
    \centering
    \includegraphics[width=.8\linewidth]{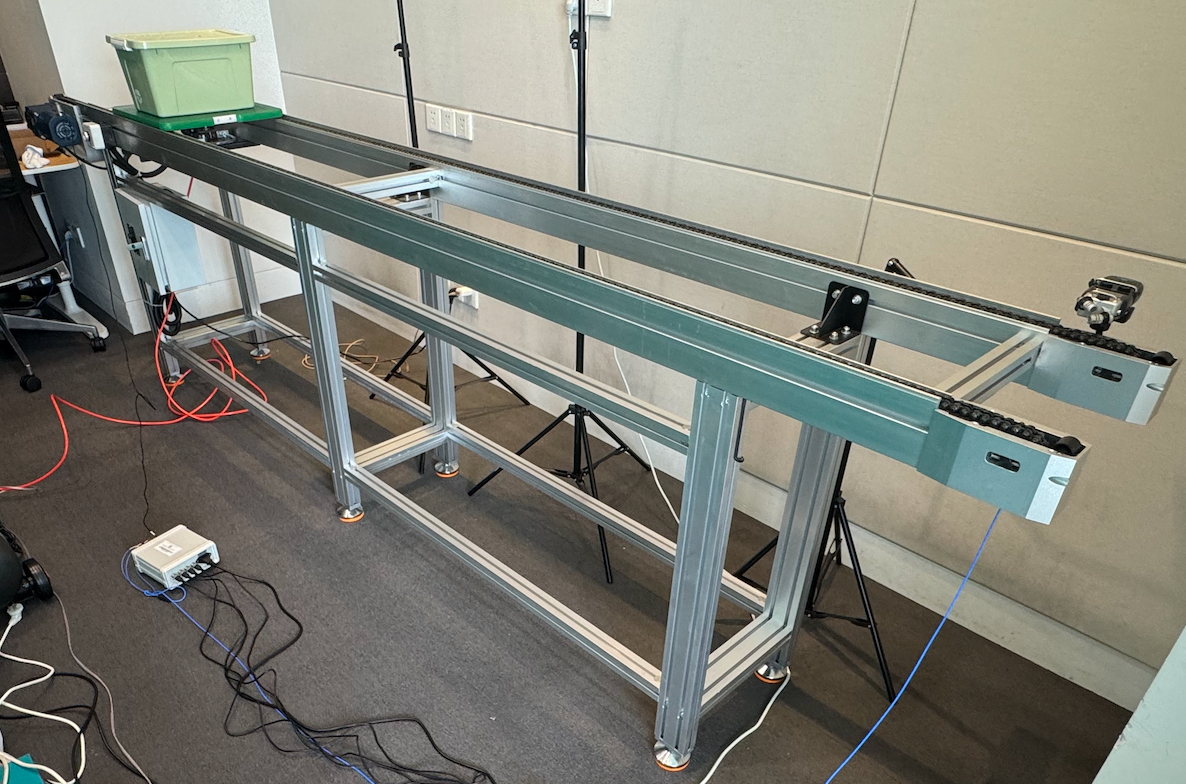}
    \vspace{-2mm}\caption{Physical setup of the single-speed chain conveyer experimental platform.}\vspace{-4mm}
    \label{fig:system_overview}
\end{figure}

The experimental platform is constructed to emulate realistic production-line operation, as illustrated in Fig.~\ref{fig:system_overview}. In practical industrial settings, production lines are typically long and linear, with unidirectional material flow rather than closed-loop circulation~\cite{deka2024comprehensivereview}. To reflect this characteristic, the experimental system is configured to prevent workpiece return by installing an air-stop mechanism which is commonly adopted in the workshop at the terminal section of the conveyor, ensuring unidirectional transportation during operation.

Data are collected under multiple controlled operating conditions by varying the conveying speed, which is discretized into five levels (\emph{20}, \emph{40}, \emph{60}, \emph{80}, and \emph{100}), and the load level, categorized as (\emph{heavy}, \emph{medium}, and \emph{light}). Here, the speed values serve as nominal indicators of different operating regimes rather than absolute physical units, while the load categories represent relative loading conditions of the SSCC system. These operating attributes are varied independently of fault conditions and appear in both normal and faulty states.

\vspace{-2mm}\subsection{Modalities and Sensors}\vspace{-1mm}

\begin{figure}[t]
    \centering
    \includegraphics[width=\linewidth]{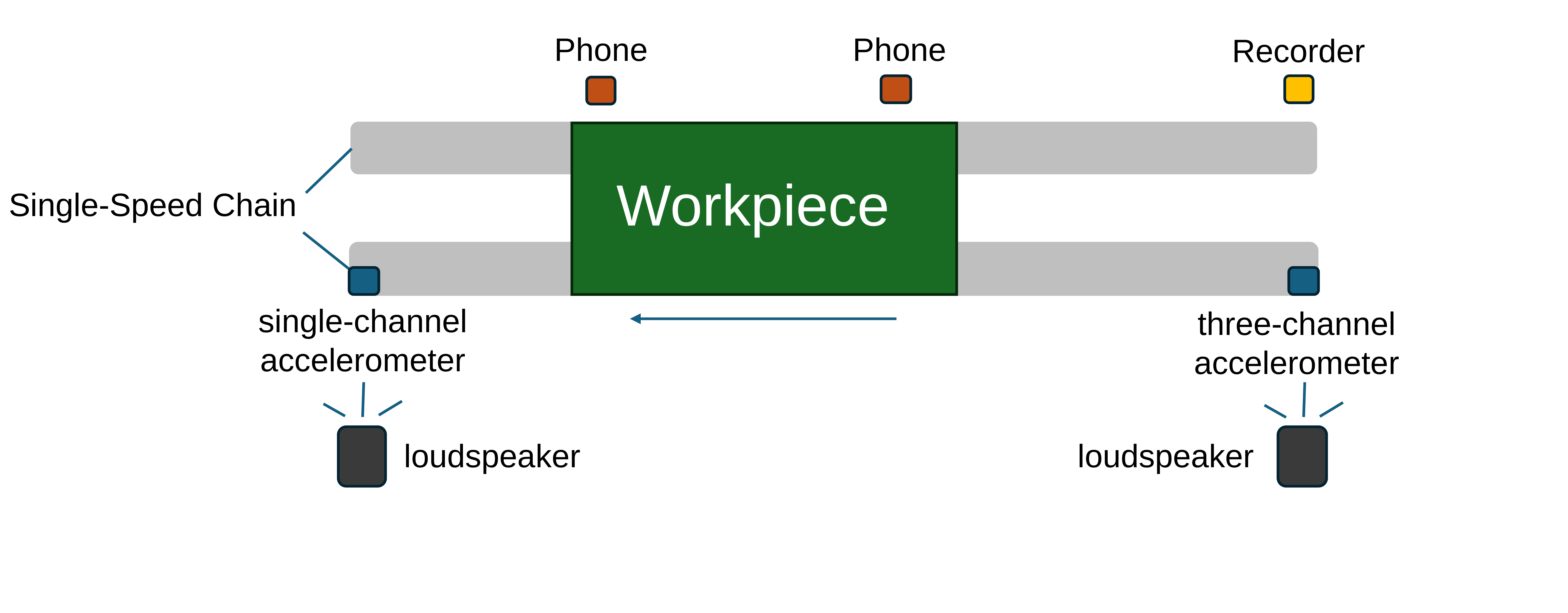}
    \vspace{-1cm}\caption{Overall system layout and sensor configuration.}\vspace{-7mm}
    \label{fig:system_layout}
\end{figure}

The SSCC dataset includes two sensing modalities, namely audio and vibration. An overview of the sensor placement and data acquisition setup is shown in Fig.~\ref{fig:system_layout}.

Audio signals are captured using three recording devices: a Zoom~H5~Handy~Recorder, an iPhone~11, and a Xiaomi~Mi~9~SE smartphone. All devices are mounted on fixed support structures and oriented toward the operating chain system to ensure consistent acoustic capture conditions. The use of multiple devices introduces natural channel diversity due to differences in microphone characteristics and recording pipelines.

In addition to audio, the mobile devices also record video during data collection. These video recordings are not used in the experiments presented in this work and are released as supplemental data to support potential future studies involving visual or audio-visual analysis.

Vibration signals are acquired using two sensors with different configurations. A single-axis vibration sensor is mounted on the motor to capture localized motor-related vibrations, while a tri-axis vibration sensor is installed at the opposite end of the system to measure system-level vibration responses along three orthogonal directions across the transmission structure.

By combining different audio devices with spatially distributed vibration sensors, all sensing modalities are synchronously recorded and treated as parallel channels, supporting research on channel-wise representation learning and multimodal fusion for industrial fault analysis.

\vspace{-2mm}\subsection{Noise Modeling and Injection}\vspace{-2mm}

Environmental noise is unavoidable in real industrial environments and has been shown to substantially degrade the performance of industrial fault analysis models~\cite{environmental_noise, gearnoisy}. Therefore, the dataset incorporates realistic environmental noise recorded in actual factory settings. During data acquisition, the recorded noise is reproduced through two loudspeakers placed around the experimental system, as illustrated in Fig.~\ref{fig:system_layout}.

For each operating condition, data are collected under both clean and noisy settings, covering normal and faulty states. This design enables systematic evaluation of noise robustness and supports reproducible benchmarking of model performance under noisy industrial conditions.

\vspace{-2mm}\subsection{Fault Types and Annotation}\vspace{-2mm}
\label{subsec:fault_type}

The dataset covers normal operation and multiple abnormal conditions that are commonly discussed in the context of chain-driven conveyor and transmission systems~\cite{wen2023faultidentification,shuai2019systemchain,wang2021response}. Specifically, four representative fault types are considered: \emph{lean}, corresponding to guide rail misalignment; \emph{dry}, indicating insufficient chain lubrication; \emph{loose}, referring to excessive looseness of the chains on both sides; and \emph{screwdrop}, corresponding to foreign-object intrusion events that may cause local obstruction or jamming in the chain transmission path.

All samples are annotated with sample-level fault labels, enabling both unsupervised fault detection and supervised fault classification studies.

\vspace{-2mm}\subsection{Dataset Organization and Signal Modalities}\vspace{-2mm}

In this dataset, a \emph{condition} is defined as a specific combination of operating attributes, including conveying speed, load level, operating category (normal or fault type), and noise setting. Each condition thus corresponds to a distinct operating scenario of the single-speed chain system.

For each condition, data are collected under steady-state operation and segmented into 5-second clips. Data acquisition is performed simultaneously across all sensing devices and are treated as synchronized channels of the same sample.

In total, the dataset contains 6{,}669 samples across all conditions. The detailed distribution of our dataset is summarized in Table~\ref{tab:dataset_overview}. For low conveying speeds (20 and 40), data are collected only under the normal and clean setting, as such speeds are uncommon in realistic industrial operating conditions.

Each data instance consists of seven signal channels recorded simultaneously from heterogeneous sensors. Specifically, three stereo audio streams are included: recorder audio stored in WAV format at 44.1~kHz, iOS audio sampled at 44.1~kHz, and Android audio sampled at 48~kHz. Although the iOS and Android recordings are stored in MP4 format, only the audio tracks are used in the modeling pipeline. In addition, vibration data are stored in the CSV format and consist of four independent channels, all sampled at 100~kHz.

\begin{table}[t]
\centering
\caption{Fault detection performance~(AUROC) of pre-trained audio encoders under different views.}\vspace{-2mm}
\label{tab:auroc_view_comparison}\small
\begin{tabular}{lccc}
\hline
\textbf{Feature} & \textbf{Audio} & \textbf{Vibration} & \textbf{Fused} \\
\hline
BEATs~\cite{chen2022beats}   & 0.868 & 0.784 & 0.891 \\
CED~\cite{dinkel2024ced}     & 0.835 & 0.772 & 0.854 \\
DaSheng~\cite{dinkel2024dasheng} & 0.760 & 0.301 & 0.681 \\
EAT~\cite{chen2024eat}     & 0.747 & 0.291 & 0.666 \\
ECHO~\cite{ECHO2025}    & 0.807 & 0.201 & 0.658 \\
FISHER~\cite{FISHER2025}  & 0.954& 0.546 & 0.882 \\
\hline
\end{tabular}\vspace{-6mm}
\end{table}

\vspace{-2mm}\section{Experimental Setup}

\subsection{Task Design and Data Partitioning}

Following common practice in industrial fault analysis, we consider two basic tasks: fault detection and classification~\cite{ECHO2025,FISHER2025}, and adopt task-specific data partitioning strategies accordingly.

Fault detection is evaluated under a normal-only training regime with a zero-shot split with respect to operating velocity. Specifically, the training set contains only normal samples collected under non-target velocities ($\mathrm{vel} \neq 100$), while the testing set consists of all normal samples collected at the target velocity ($\mathrm{vel}=100$) across different load and noise conditions, together with all abnormal samples from all operating regimes.

Fault classification is formulated as a supervised multi-class classification task. To increase the difficulty of generalization, all samples collected at velocity 80 are fully held out from training. In addition, samples belonging to selected fault categories (dry and lean) under velocity 100 are removed from the training set and reassigned to the testing set.

Data partitioning is performed at the condition level. All signal segments belonging to the same condition are assigned exclusively to either training or testing, preventing data leakage.

\begin{table*}[t]
\centering
\caption{Fault classification performance of pre-trained audio encoders under different views. Acc denotes overall accuracy, BalAcc denotes the mean recall across classes, and MacroF1 denotes the mean F1 score across classes.}\vspace{-2mm}
\label{tab:fault_classification_views}
\resizebox{.8\textwidth}{!}{\large
\begin{tabular}{llcccccccc}
\hline
Feature & View & Acc & BalAcc & MacroF1 & F1(dry) & F1(lean) & F1(loose) & F1(normal) & F1(screwdrop) \\
\hline
BEATs~\cite{chen2022beats} & Audio     & 0.914 & 0.900 & 0.906 & 0.954 & 0.935 & 0.908 & 0.959 & 0.774 \\
      & Vibration & 0.869 & 0.873 & 0.884 & 0.985 & 0.984 & 0.803 & 0.839 & 0.811 \\
      & Fused     & 0.941 & 0.928 & 0.937 & 0.996 & 0.996 & 0.903 & 0.993 & 0.796 \\
\hline
CED~\cite{dinkel2024ced}   & Audio     & 0.817 & 0.777 & 0.787 & 0.846 & 0.760 & 0.876 & 0.861 & 0.592 \\
      & Vibration & 0.843 & 0.842 & 0.849 & 0.996 & 0.998 & 0.753 & 0.891 & 0.609 \\
      & Fused     & 0.905 & 0.893 & 0.902 & 0.995 & 0.997 & 0.849 & 0.963 & 0.705 \\
\hline
DaSheng~\cite{dinkel2024dasheng} & Audio     & 0.924 & 0.904 & 0.913 & 0.922 & 0.864 & 0.956 & 0.951 & 0.870 \\
        & Vibration & 0.910 & 0.914 & 0.922 & 0.999 & 0.999 & 0.861 & 0.881 & 0.869 \\
        & Fused     & 0.967 & 0.959 & 0.965 & 0.997 & 0.997 & 0.950 & 0.989 & 0.891 \\
\hline
EAT~\cite{chen2024eat}   & Audio     & 0.850 & 0.831 & 0.832 & 0.832 & 0.776 & 0.918 & 0.898 & 0.733 \\
      & Vibration & 0.871 & 0.873 & 0.883 & 0.960 & 0.961 & 0.808 & 0.860 & 0.826 \\
      & Fused     & 0.927 & 0.916 & 0.924 & 0.962 & 0.962 & 0.895 & 0.973 & 0.830 \\
\hline
ECHO~\cite{ECHO2025}  & Audio     & 0.975 & 0.971 & 0.973 & 0.963 & 0.973 & 0.983 & 0.983 & 0.963 \\
      & Vibration & 0.715 & 0.690 & 0.682 & 0.707 & 0.809 & 0.720 & 0.784 & 0.391 \\
      & Fused     & 0.907 & 0.886 & 0.894 & 0.954 & 0.963 & 0.871 & 0.990 & 0.693 \\
\hline
FISHER~\cite{FISHER2025} & Audio     & 0.969 & 0.969 & 0.968 & 0.970 & 0.951 & 0.970 & 0.977 & 0.970 \\
       & Vibration & 0.886 & 0.886 & 0.890 & 0.991 & 0.994 & 0.801 & 0.951 & 0.710 \\
       & Fused     & 0.955 & 0.946 & 0.953 & 0.997 & 0.997 & 0.923 & 0.998 & 0.849 \\
\hline
\end{tabular}
}\vspace{-6mm}
\end{table*}

\begin{figure}[!t]
    \centering
    \includegraphics[width=.5\textwidth]{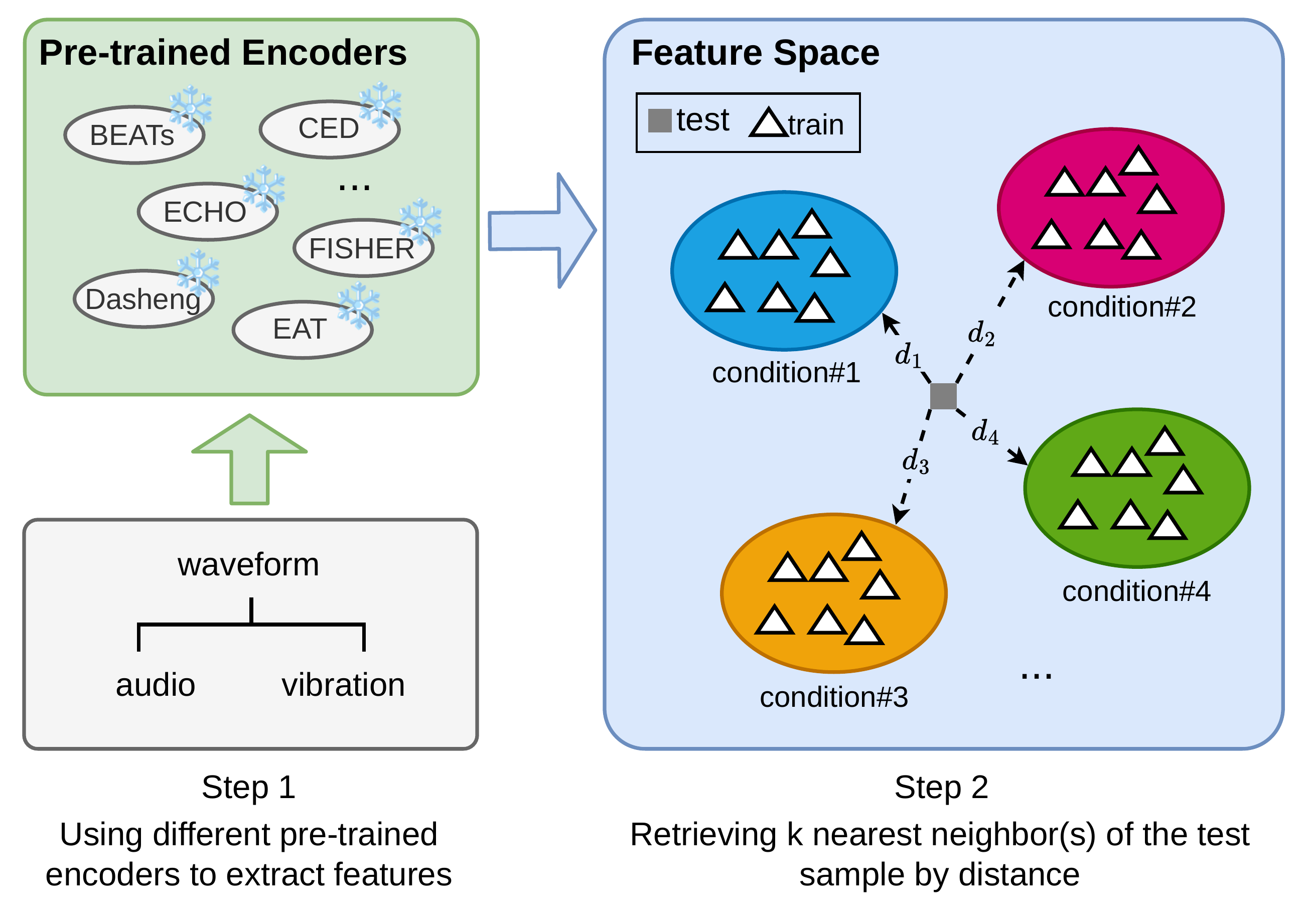}
    \vspace{-8mm}\caption{An evaluation pipeline on the SSCC dataset. For AD task, the distance to the nearest neighbor is used as the anomaly score. For FD task, the prediction is determined by majority vote among k nearest neighbors' labels.}
    \label{fig:eval_pipeline}\vspace{-6mm}
\end{figure}

\vspace{-2mm}\subsection{Evaluation Protocols}

To ensure a fair and unified evaluation across different tasks and datasets, we adopt
a kNN-based inference framework as the benchmark method for fault detection
tasks~\cite{jiang2024anopatch}.
This choice is able to assess the intrinsic quality of extracted embeddings
without introducing additional task-specific model bias, which is a common evaluation method for testing the capability of the encoders~\cite{xares}.

As shown in Figure~\ref{fig:eval_pipeline}, all models are evaluated under the same kNN-based pipeline, with feature embeddings
extracted using the open-source SIREN toolkit~\cite{ECHO2025}, which provides a unified
interface for feature extraction with different pretrained models.
The embeddings from normal datasets form a memory bank.
During inference, each test embedding is compared against the memory bank using cosine distance.
Unless otherwise specified, the number of nearest neighbors is fixed to $k=1$ for fault detection and $k=11$ for fault classification, as this choice provides stable and consistent performance across different settings while balancing the bias-variance trade-off. No dataset- or model-specific parameter tuning is performed.

Both fault detection and fault classification follow a unified channel-wise kNN inference
framework across seven channels. For each channel, features are standardized using a channel-specific standard scaler fitted on training normal samples only.

To compare the representation quality of different encoders, we include six pre-trained foundation models for comparison:
BEATs~\cite{chen2022beats}, CED~\cite{dinkel2024ced}, EAT~\cite{chen2024eat},
Dasheng~\cite{dinkel2024dasheng}, FISHER~\cite{FISHER2025}, and
ECHO~\cite{ECHO2025}. All models apply the ViT-style~\cite{dosovitskiy2020image}
architecture as the backbone and are trained on large-scale open-source audio datasets
across diverse domains.

For each test sample, the Euclidean distance to the $k$-th nearest neighbor is computed
channel-wise and used as the anomaly measure. To ensure cross-channel comparability, distances are normalized via empirical cumulative distribution function (ECDF)–based quantile mapping~\cite{li2022ecod,ge2025dccopgan}, yielding channel-wise anomaly scores in $(0,1)$. Final anomaly scores are obtained by late fusion through averaging, and performance is evaluated using the area under the receiver operating characteristic curve (AUROC), which is popular metric on the fault detection task~\cite{DCASE2025}. For fault classification, the $k$ nearest neighbors are retrieved per channel, and class probabilities are computed via distance-weighted voting. Channel-wise probability vectors are aggregated through late fusion, and the final label is obtained by $\arg\max$. Performance is reported using accuracy, balanced accuracy, and macro-F1 score.

\vspace{-2mm}\section{Results and Analysis}\vspace{-2mm}
After feature extraction and downstream fault detection and calssification tasks, the results are shown in Table~\ref{tab:auroc_view_comparison} and Table~\ref{tab:fault_classification_views}.

For fault detection, audio-based methods generally outperform vibration-based ones across most feature extractors, while vibration achieves slightly higher performance for BEATs and CED. Under the proposed normal-only training and cross-velocity evaluation protocol, this indicates that acoustic signals provide more discriminative cues for abnormal conditions in the studied system. A plausible explanation is that common fault types introduce subtle sound variations that are more readily captured by audio representations, whereas vibration signals are often dominated by strong periodic components of steady-state operation. Decision-level fusion generally improves performance over vibration-only views and in some cases matches or exceeds audio-only results, demonstrating the benefit of combining audio and vibration signals.

Fault classification achieves high accuracy across all evaluated feature extractors. Under this task, audio and vibration exhibit complementary strengths: vibration-based features perform better for the \emph{screwdrop} fault, likely due to impulsive mechanical responses, while audio-based features consistently achieve higher F1 scores for the \emph{loose} fault, which is associated with frictional and rattling sounds. As a result of these complementary characteristics, the fused multimodal system achieves higher performance than single-modality ones in most cases, except for ECHO and FISHER, indicating the advantage of integrating audio and vibration signals for fault classification. 

Overall, these results reveal a clear task-dependent difference in modality effectiveness within the dataset. Audio signals are more informative for fault detection under the unsupervised evaluation protocol, whereas fault classification with explicit labels benefits from the complementary physical cues captured by both audio and vibration. This indicates that modality contribution depends jointly on the target task and fault characteristics. While simple distance-based baselines already provide reasonable performance, the dataset offers substantial headroom for future exploration of more advanced acoustic representation learning and multimodal fusion approaches.

\vspace{-2mm}\section{Conclusion}\vspace{-1mm}

This work introduces a multimodal industrial fault analysis dataset with seven synchronized sensing channels, capturing representative fault conditions observed in SSCC system. By jointly providing audio and vibration measurements across multiple operating regimes, including varying speeds, load levels, and environmental noise, the dataset enables systematic and reproducible evaluation of multimodal approaches for speed chain conveyor fault analysis. Using a unified channel-wise kNN evaluation probe, we study fault detection and classification under task-specific protocols and different sensing modalities capture complementary aspects of fault behavior. These findings suggest that the dataset contains rich multimodal information that has not yet been fully exploited by current encoders, leaving substantial potential for further performance gains. The dataset therefore serves as a practical and extensible benchmark for multimodal industrial fault analysis, providing a solid foundation for future research on more advanced representation learning and fusion strategies.
\FloatBarrier

\pagebreak

\section{Generative AI Usage Disclosure}
Generative AI tools were used solely for grammar correction and language polishing in this manuscript. All research content, analyses, and conclusions were independently developed by the authors.

\bibliographystyle{IEEEtran}
\bibliography{ref}

\end{document}